# Field-temperature phase diagram of magnetic bubbles spanning charge/orbital ordered and metallic phases in La$_{1-x}$Sr$_x$MnO$_3$ ($x$ = 0.125)

A. Kotani[1], H. Nakajima[1], K. Harada[1,2], Y. Ishii[1] and S. Mori[1,*]

[1]Department of Materials Science, Osaka Prefecture University, Sakai, Osaka 599-8531, Japan.
[2]Center for Emergent Matter Science, the Institute of Physical and Chemical Research (RIKEN), Hatoyama, Saitama 350-0395, Japan.

**Abstract**

We report formation of magnetic textures in the ferromagnetic (FM) phase of La$_{1-x}$Sr$_x$MnO$_3$ for $x$ = 0.125; these textures are magnetic bubbles, magnetic stripe domains, and forced FM states. *In situ* Lorentz microscopy (LM) observations show that magnetic bubbles exist in the FM insulating phase accompanying the formation of the charge/orbital ordering (CO/OO). Furthermore, stable magnetic bubbles still exist in an intermediate temperature region between the CO/OO ($T_{CO}$=155 K) and FM ($T_c$ =190 K) transition temperatures. These magnetic bubbles are believed to originate from the magnetocrystalline anisotropy and the dipole–dipole interaction in the FM phase. Based on *in situ* LM observations as a function of both temperature and the strength of the external magnetic field applied, a magnetic field–temperature phase diagram is constructed, exhibiting the stabilizing regions of the magnetic bubbles in the FM phase of La$_{0.875}$Sr$_{0.125}$MnO$_3$.

**Introduction**

Numerous nanoscale magnetic textures have recently been investigated with respect to applying them to high-performance spin devices; magnetic skyrmions, for example, have been found to possess vortex-like nanometric spin structures, and their spin swirling is defined by an integer winding number [1-5]. Studies have shown that magnetic skyrmions are induced by the Dzyaloshinskii–Moriya (DM) interaction in chiral magnets that have a $B$20-type cubic structure without inversion symmetry; such magnets include MnSi and FeGe [6-10]. Magnetic bubbles in uniaxial ferromagnets that have inversion symmetry have also been found to have the same spin structure as that of the skyrmions [11]. Magnetic bubbles are induced by applying external magnetic fields parallel to a magnetic easy axis, and they are stabilized by magnetic anisotropy [12-14]. Additionally, magnetic dipole–dipole interactions stabilize the swirling-spin configuration in a magnetic bubble's domain wall. These magnetic bubbles are called type-I magnetic bubbles, and they have the same winding number as the skyrmions. Therefore, the type-I magnetic bubbles have topologically non-trivial spin structure with a finite topological charge. On the other hand, the magnetic bubbles called type-II magnetic bubbles without spin rotation have topologically trivial structures. A large number of magnetic skyrmions in metallic magnets have been studied [4-10,15-18], but few studies have been conducted on magnetic skyrmions in the insulating phase; examples of materials studied include $Cu_2OSeO_3$ and $GaV_4S_8$ [19,20, 21].

Hole-doped $La_{1-x}Sr_xMnO_3$ has been known to exhibit various magnetic phases and crystal structures [22-24]. In addition, many types of magnetic textures such as magnetic bubbles and magnetic stripe domains have been found in the ferromagnetic (FM) phase of $La_{1-x}Sr_xMnO_3$ [25-28]. When $x = 0.125$, $La_{1-x}Sr_xMnO_3$ exhibits two successive structural phase transitions; one is a transition from an orthorhombic structure that has a $P$bnm symmetry ($O$, $c/\sqrt{2} < b \approx a$) to a monoclinic structure that has a $P12_1/c1$ symmetry ($M'$, $a/\sqrt{2} < b \approx c$) in the paramagnetic insulating (PM-I) phase accompanying the Jahn–Teller (JT) distortion of the $Mn^{3+}$ ion. The other is a transition from the monoclinic structure ($M'$) in the ferromagnetic metallic (FM-M) state below $T_C \approx 190$ K to the monoclinic structure ($M''$) with a charge/orbital-ordered (CO/OO) structure in the ferromagnetic insulating (FM-I) state at $T_{CO} \approx 150$ K [29]. The CO/OO structure here is characterized as a modulated structure with $q$ = <1/2 1/4 1/4> [30]. Li *et al.*, however, found that the spin orientation in the FM-I phase at 5 K is almost parallel to the [110] direction, where, for simplicity, the notations are based on the high-temperature cubic structure with a $Pm\bar{3}m$ space group (SG). Thus, in the $M''$ structure, the magnetic easy axis keeps almost parallel to the [110] direction in the FM-I phase [29].

In a previous study, we investigated magnetic bubbles in the FM-M phase of $La_{1-x}Sr_xMnO_3$ for $x = 0.175$, which had an orthorhombic structure with inversion symmetry; we used *in situ* Lorentz microscopy (LM) observations combined with small-angle electron diffraction experiments [26,27]. We found that magnetic textures such as magnetic bubbles and magnetic stripe domains appeared in $La_{0.825}Sr_{0.175}MnO_3$, and they depended significantly on the magnetocrystalline anisotropy. Recently, Nagai *et al.* examined the magnetic domain structures in the CO/OO FM-I phase of $La_{0.875}Sr_{0.125}MnO_3$ using LM observations, and they found that magnetic stripe domains transform into elliptically shaped magnetic nanodomains when an external magnetic field is applied perpendicular to a thin film [31]. Notably, in the LM observations conducted by Nagai *et al.*, the orientation of the magnetization was not parallel to the magnetic easy axis and consequently the elliptical magnetic domains were formed by application of magnetic field along the [111] direction. From the research currently available, it is unclear whether magnetic bubbles form owing to the application of an external magnetic field parallel to the magnetic easy axis in the FM-I phase of $La_{0.875}Sr_{0.125}MnO_3$.

In this present work, we investigated the magnetic textures in the FM phase of $La_{0.875}Sr_{0.125}MnO_3$ as both a function of temperature and the strength of external magnetic fields; this was done through the use of *in situ* LM observations and small-angle electron diffraction experiments. In particular, we investigated different magnetic textures by applying an external magnetic field along the magnetic easy axis in the FM phase of $La_{0.875}Sr_{0.125}MnO_3$. Our results show that in the FM-I phase that has a CO/OO structure, type-I magnetic bubbles with clockwise (CW) and counter clockwise (CCW) spin rotations exist and they are stable. Using LM and a phase reconstruction technique based on a transport-of-intensity equation (TIE) [32, 33], we found the spin configuration of the type-I magnetic bubbles. Because of our experimental results, we were able to construct a magnetic field–temperature phase diagram in the FM phase of $La_{0.875}Sr_{0.125}MnO_3$.

**Experimental method**

Single crystalline samples of $La_{0.875}Sr_{0.125}MnO_3$ were prepared using a conventional solid-state reaction and the floating zone method [22]. The crystal structures were determined by powder X-ray diffraction. Note that the magnetic easy axis of $La_{0.875}Sr_{0.125}MnO_3$ is parallel to the [110] direction. We characterized the crystalline orientations in the obtained single crystals and determined them by back-plate Laue-type X-ray diffraction experiments. The electrical resistivity was measured using a Physical Property Measurement System (PPMS) (Quantum Design, Inc.) that used a conventional four-terminal method. The magnetization was measured using a vibrating sample magnetometer (VSM) equipped with PPMS. The thin films for the LM observations were prepared by $Ar^+$-ion sputtering at room temperature. Our *in situ* LM observations were conducted by a TEM (JEM-2100F) that used a double-tilted cooling holder in the temperature region of 80–298 K. External magnetic fields were applied to the thin films by exciting the objective lens. For simplicity, all indices of the crystal planes and orientations were indexed based on the cubic perovskite structure with $Pm\bar{3}m$ SG.

**Experimental results**

First, to confirm the structural and physical properties of the single crystals of $La_{0.875}Sr_{0.125}MnO_3$ that were obtained, we investigated the electron diffraction patterns and the crystals' electrical and magnetic properties. In the temperature ($T$) dependence of the electrical resistivity ($\rho$), as shown in Fig. 1(a), there is a kink at approximately 260 K; this indicates that there is a structural phase transition from the orthorhombic to the monoclinic structure. The increase in $\rho$ at 190 K can be ascribed to a ferromagnetic transition, and the decrease at 155 K corresponds to the metal–insulator transition that occurs when the CO/OO structure forms. These results are consistent with those obtained by a previous study [34]. To clarify both the structural phase transitions and the formation of the CO/OO structure, we investigated how the electron diffraction patterns varied with temperature between 80 and 298 K. In the high-temperature paramagnetic phase, fundamental reflection spots due to the orthorhombic structure with *P*bnm SG can be observed, as shown in the top panel of Fig. 1(b). As the temperature fell below 170 K, superlattice reflection spots appeared at the *h, k,* and *l* + 1/2 reciprocal positions in the electron diffraction pattern with the [110] incidence; this implies that a structural phase transition to the monoclinic structure with $P12_1/c1$ SG took place (note that *h, k,* and *l* are all integers). As the temperature decreased to below $T_{CO}$ = 155 K, superlattice reflection spots were observed at the *h, k,* and *l* + 1/4 reciprocal positions, as indicated by the yellow arrows in the lower panel of Fig. 1(b); these indicate the formation of the CO/OO structure. These results are consistent with those of a previous study that used electron diffraction measurement techniques [30,35]. In this earlier study, the $M'$ structure in the intermediate temperature region ($T_s > T > T_{CO}$) was characterized by the presence of a large JT distortion, while the $M''$ structure in the lower temperature region (*i.e.*, below $T_{CO}$) was characterized by the formation of the CO/OO structure. Revealing the magnetic anisotropy in the two monoclinic structures ($M'$ and $M''$) is essential to understand the conditions wherein magnetic bubbles are formed. Because of this, we examined different magnetic anisotropy constants ($K_u$) as a function of temperature in the FM phase of $La_{0.875}Sr_{0.125}MnO_3$. Figure 1(c) shows the changes in the magnetization curves as a function of temperature in the FM phase of $La_{0.875}Sr_{0.125}MnO_3$. Magnetic fields were applied along both the [110] axis and the crystal plane perpendicular to the [110] axis. The values of the saturated magnetization at $T$ = 170, 120, and 80 K were approximately 3.12, 3.55, and 3.75 $\mu_B$, respectively. The anisotropic magnetic field ($H_k$) in the FM phase was obtained from the magnetization curves at $T$ = 170, 120, and 80 K. $H_k$ is defined as being the critical value of the magnetic field above which the difference in magnetization between $H // c$ and $H // ab$ is less than 2%. The values of $H_k$ at $T$ = 170, 120, and 80 K are 0.95, 0.77, and 0.76 T, respectively. The $K_u$ values obtained from the anisotropic magnetic fields of $T$ = 170, 120, and 80 K were 2.31, 2.13, and 2.22 × 10$^5$ J/m$^3$, respectively. These values satisfied the formation conditions of the magnetic bubbles [12,13]. This suggests that magnetic bubbles can be formed in the FM phase of $La_{0.875}Sr_{0.125}MnO_3$ when an external magnetic field is applied perpendicular to the thin films in a manner that would make them parallel to the magnetic easy axis along the [110] direction.

We investigated the magnetic textures in the FM phase of $La_{0.875}Sr_{0.125}MnO_3$ by using *in situ* LM observations. We found that magnetic stripe domains appear in the CO/OO FM-I phase at 90 K, as shown in Fig. 2(a). The residual magnetic field in the objective lens of the TEM was observed to be less than 20 mT. The spatial configuration of the spins in the magnetic stripe domains was observed to be the same as that in the FM phase of $La_{0.825}Sr_{0.175}MnO_3$ [26]. A significant feature observed in the domains was the formation of zigzag walls, as shown in the inset of Fig. 2(a). External magnetic fields were applied perpendicular to the thin films along the [110] magnetic easy axis by exciting the objective lens of the TEM. As the strength of the external magnetic fields increased to 340 mT, the magnetic stripe domains changed into type-I magnetic bubbles with a diameter of 150 nm (Fig. 2(b)). Type-I and type-II magnetic bubbles were found to co-exist in a random manner at 420 mT, and this can be clearly seen in Fig. 2(c). The spatial configurations of the spins inside the type-I and type-II magnetic bubbles are schematically described in Fig. 2(f) [27]. The presence of these two types of magnetic bubbles is entirely distinct from the magnetic nanodomains found by Nagai *et al*. [31]; for the circular-shaped magnetic bubble found in this work, spins rotates in the magnetic domain wall and are oriented along the [110] direction, which is clearly contrary to spins simply directed along the [11$\bar{1}$] direction in the elliptical magnetic domain in the previous work by Nagai *et al*.. This implies that applying external magnetic fields along the [110] magnetic easy axis is essential for inducing magnetic bubbles. As the strength of the external magnetic fields was increased further, the size of the bubbles began to decrease, shrinking from 150 to 100 nm at 420 mT before becoming approximately 80 nm at 470 mT (Fig. 2(d)). At 530 mT, the magnetic bubbles disappeared completely and the forced ferromagnetic state was realized, as shown in Fig. 2(e).

To clarify the spatial configuration of the spins in the type-I magnetic bubbles in Fig. 2(c), we analyzed the spin configurations of the bubbles indicated by the white squares A and B in Fig. 3(a). A phase retrieval technique based on the transport-of-intensity equation (TIE) was used in this analysis. The spatial distribution of the type-I magnetic bubbles that had dark and bright contrasts at their centers stemmed from the clockwise (CW) and counter clockwise (CCW) spin rotations at their respective domain walls. We analyzed the LM images of the magnetic bubbles (Figs. 3(b) and 3(c)) using the TIE analysis method, and with this we were able to reproduce the orientation and magnitude of the in-plane magnetization. We found that the distinctive feature of the bubbles in Fig. 3(c) was that they had no bright or dark points at their core, unlike the type-I magnetic bubbles found in $La_{0.825}Sr_{0.175}MnO_3$ [26-28]. For comparison, the magnetic bubble formed in $La_{0.825}Sr_{0.175}MnO_3$ is shown in Fig. 3(d), [27]. In Fig. 3(e), a schematic illustration of the transition process from the magnetic stripe domains to the type-I magnetic bubbles caused by the application of external magnetic fields is given, and we found that it is similar to the formation process of the type-I magnetic bubbles of $La_{0.825}Sr_{0.175}MnO_3$ [26,27].

As shown in Fig. 3, we can see that type-I magnetic bubbles are formed in the CO/OO FM-I phase. We can see that $La_{0.875}Sr_{0.125}MnO_3$ exhibits an insulating-to-metallic transition in the FM phase at approximately 150 K; as such, we focused on whether magnetic bubbles are formed in the FM-M phase between 150–195 K. Figures 4(a)–(c) show a series of magnetic textures obtained at 170 K as a function of the strength of the external magnetic fields; magnetic stripe domains can be observed at 20 mT. The magnetic stripe domains in Fig. 4(a) form elliptical domains at 215 mT, as indicated by the white broken lines in Fig. 4(b) and the contour highlighted by the yellow broken lines in the inset of Fig. 4(b); meanwhile, no magnetic contrast is observed at 250 mT, as shown in Fig. 4 (c). Figures 4 (d)–(f), however, show that the magnetic bubbles vary as a function of temperature in a constant 350 mT magnetic field; type-I magnetic bubbles are formed at 100 K and 350 mT, as shown in Fig. 4(d). As the temperature was increased in this constant magnetic field, the magnetic bubbles shrank from 150 nm to 100 nm at 130 K and disappeared altogether at 160 K. These results suggest that magnetic bubbles appear in the narrow range of 200–250 mT.

We also investigated how the magnetic stripe domains vary as a function of temperature in a constant magnetic field of 200 mT. The magnetic stripe domains at 130 K in Fig. 5(a) start to transform into magnetic bubbles at 150 K (Fig. 5(b)), and the domains and bubbles co-exist at 170 K (Fig. 5(c)). No magnetic contrast can be observed above the FM transition temperature of $T_C$ =195 K. Our findings imply that magnetic bubbles exist and are stable in the FM-M phase below $T_C$ = 195 K.

Based on the *in situ* LM observations we made, we constructed a *T–B* phase diagram with respect to the stable magnetic bubbles, as shown in Fig. 6. The circles in the phase diagram show the experimental points determined by the *in situ* LM observations. A significant feature in the phase diagram is that the existence of the CO/OO structure has some influence on the formation of the magnetic bubbles, as discussed below. In addition, in the temperature range between 155–195 K in the FM-M phase, the magnetic bubbles exist stably in the magnetic field region below 200 mT, and they exist stably in a magnetic field of 100 mT just below the FM transition temperature of $T_C$ =195 K.

**Discussion**

In the inset of Fig. 2(a), we can see zigzag magnetic stripe domains in the walls of the CO/OO FM-I phase of $La_{0.875}Sr_{0.125}MnO_3$. In our previous work on the phase-separated manganite, $La_{0.25}Pr_{0.375}Ca_{0.375}MnO_3$ [36], FM domains with zigzag domain walls were found to be formed in the phase-separated state consisting of the CO/OO and FM-M states. In the case of $La_{0.25}Pr_{0.375}Ca_{0.375}MnO_3$, the nucleation of the FM-M domains inside the PM-CO/OO phase occurred as the temperature decreased below $T_C$ = 50 K and zigzag magnetic domain walls separated the FM domains and the CO-OO domains [36]. We believe that the pinning force originating from the structural differences between the crystal structures in the FM CO/OO phases had a pronounced influence on the phase boundary. However, the magnetic stripe domain walls in $La_{0.875}Sr_{0.125}MnO_3$ are characterized by zigzag magnetic domain walls, as shown in the inset of Fig. 2(a). As reported by Asaka *et al.*, a CO/OO structure with a nanometric coherent length of approximately 10 nm was superimposed on the FM state [35]. It is suggested that the JT distortion in the CO/OO structure had some significant influence on the formation of the magnetic stripe domains, which resulted in the zigzag magnetic domain walls. As such, a local strain effect due to the JT distortion should originate in the zigzag magnetic stripe domain walls.

The formation of type-I magnetic bubbles without bright or dark points at their center is found in $La_{0.875}Sr_{0.125}MnO_3$, as shown in Fig. 3(a). As we have previously reported, type-I magnetic bubbles possess a bright point at their core in the FM-M phase of $La_{0.825}Sr_{0.175}MnO_3$ at 90 K [26-28] in Fig. 3(d). The bright point originates from the convergence of the transmission electron beam due to the rotation of the magnetization around the core of a magnetic bubble. However, the magnetic bubbles in the FM-I phase of $La_{0.875}Sr_{0.125}MnO_3$ have no bright point; this implies that the spatial configuration of the spins inside the magnetic bubbles of $La_{0.875}Sr_{0.125}MnO_3$ and $La_{0.825}Sr_{0.175}MnO_3$ are different. Consequently, we considered the magnetocrystalline anisotropy in the FM phase of $La_{0.875}Sr_{0.125}MnO_3$ based on the anisotropic field ($H_k$) obtained in Fig. 1(a). The values of the anisotropic field, $H_k$ [T], and saturation magnetization, $M_s$ [$\mu_B$ / f.u.], obtained by magnetization at 8 T were 0.95 and 3.12 (at 170 K), 0.77 and 3.55 (at 120 K), and 0.76 and 3.75 (at 80 K). The values of the magnetocrystalline anisotropy constants ($K_u$) calculated by using the equation $K_u = (1/2) H_k \times M_s$ at each temperature in $La_{0.875}Sr_{0.125}MnO_3$ were 2.22 (at 80 K), 2.13 (at 120 K), and $2.31 \times 10^5$ J/m$^3$ (at 170 K). The magnetocrystalline anisotropy in the FM phase has a constant value of approximately $2.2 \times 10^5$ J/m$^3$, as while $H_k$ increases with increasing temperature, $M_s$ decreases; the effects of this are relatively greater than when $1.52 \times 10^5$ J/m$^3$ obtained in the FM phase of $La_{0.825}Sr_{0.175}MnO_3$ at 90 K [27]. Note that in the LM observations we made for the magnetic easy axis along the [110] direction, the external magnetic field to the thin films is applied parallel to the orientation of the magnetocrystalline anisotropy. This situation implies that the magnetization in the sample is oriented along the perpendicular direction to the thin films. The FM phase of $La_{0.875}Sr_{0.125}MnO_3$ exhibits greater magnetocrystalline anisotropy than that of $La_{0.825}Sr_{0.175}MnO_3$ and, as a result, the spins inside the magnetic bubbles are oriented so as to be perpendicular to the thin film; this gives rise to no bright/dark points at the core of the magnetic bubbles. In conclusion, the spatial configuration of the spins inside a magnetic bubble depends strongly on the magnetocrystalline anisotropy.

In addition, magnetic bubbles are stabilized by the shape anisotropy interacting with the demagnetizing effect except for the uni-axial magnetic anisotropy including the magnetocrystalline anisotropy along the direction perpendicular to a thin film. Therefore, magnetic bubbles are formed in the thin film as the TEM samples we observed with the thickness less than 200 nm. In bulk samples, the shape anisotropy is reduced so that the magnetic bubble state is not stable. Skyrmions induced by DM interactions also requires the shape anisotropy. Our observation revealed that magnetic bubbles were formed in the thin film of $La_{0.875}Sr_{0.125}MnO_3$. However, it has been unclear whether they are formed in the bulk sample.

**Summary**

Magnetic textures, including magnetic bubbles, in the FM phase of $La_{0.875}Sr_{0.125}MnO_3$ were investigated by means of *in situ* LM observations as functions of both temperature and the strength of external magnetic fields. We found that magnetic stripe domains transformed into magnetic bubbles in the CO/OO FM-I phase. In the magnetic bubbles, the spins rotate at the domain walls and are oriented perpendicularly to the thin films inside the magnetic bubbles. The spatial configuration of the spins inside the magnetic bubbles depends significantly on both the strength of the magnetocrystalline anisotropy and the dipole–dipole interaction in the FM phase. Based on our *in situ* LM observations, we constructed a magnetic field–temperature phase diagram for the FM phase of $La_{0.875}Sr_{0.125}MnO_3$.

This is the first observation of magnetic bubbles for a FM-I phase that has a CO/OO structure. Stable magnetic bubbles co-exist with the CO/OO structure in the specific region of the phase diagram. Magnetic bubbles with 100–200 nm in diameter in the insulating phase exhibit great potential for possible future applications of magnetic bubbles, *e.g.*, low-energy spintronic devices with unique magnetoelectric couplings could be developed using them. Moreover, $La_{0.875}Sr_{0.125}MnO_3$ exhibits the homogeneous CO/OO state in the FM phase. On the other hand, it has been recognized that perovskite manganese oxides such as $(La,Pr)_{5/8}Ca_{3/8}MnO_3$ exhibit the electronic phase separation between the FM-M and the CO/OO-I phases [36]. It is anticipated that our experimental results in the present work will help to understand the formation mechanism of magnetic bubbles in a variety of electronic phases in manganese oxides, which were produced by strong interactions among the spin, orbital and charge degrees of freedom of electrons. In addition, magnetic textures including magnetic bubbles and magnetic skyrmions in functional materials should play some crucial roles to give rise to new functionalities such as topological Hall effects.

**ACKNOWLEDGEMENTS**

This work was partially supported by Grant-in-Aid (No. 16H03833 and No. 15K13306) from the Ministry of Education, Culture, Sports, Science and Technology (MEXT), Japan.

References

* mori@mtr.osakafu-u.ac.jp

Figure captions;

Figure 1 (color online). (a) Variation of resistivity as a function of temperature in $La_{0.875}Sr_{0.125}MnO_3$. 'PM-I', 'FM-M', and 'FM-CO/OO-I' represent the magnetic phase, such as, the paramagnetic insulating, ferromagnetic metallic, and ferromagnetic charge-orbital ordered insulating phases, respectively. (b) [110]-zone axis electron diffraction patterns at (upper) 293 K, (middle) 170 K, and (lower) 120 K, showing the appearance of superlattice reflection spots at the $h$, $k,$ and $l + 1/2$ (white arrows) and $h$, $k$, and $l + 1/4$ (yellow arrowheads) reciprocal positions. (c) Variation of the $M$–$H$ hysteresis curves as a function of temperature at 170, 120, and 80 K. In the figure, 'O', 'M′', and 'M″' represent the crystal structure, as the orthorhombic, monoclinic without CO/OO, and monoclinic structure with CO/OO.

Figure 2 (color online). Variation of magntic textures at 90 K in the (110) plane as a function of the strength of the external magnetic field between 20–530 mT. The strength of the magnetic fields are (a) 20, (b) 340, (c) 420, (d) 470, and (e) 530 mT, respectively. The inset of (a) shows a magnified image of the zigzag magnetic stripe domain walls, indicated by the white broken squares in (a). The bar in the inset of (a) indicates 200 nm. (f) Schematic description of type-I and type-II magnetic bubbles. In (c), type-I and type-II magnetic bubbles are indicated by the broken and solid squares, respectively.

Figure 3 (color online). Spatial configurations of the spins inside the magnetic bubbles. (a) A Lorentz micrograph exhibiting the presence of magnetic bubbles with CW and CCW spins obtained at 90 K and $B$ = 350 mT. (b) and (c) Magnified images of two magnetic bubbles highlighted by broken squares in (a). Color representations of the in-plane magnetization are calculated with the aid of the TIE analysis method.

The directions of the in-plane magnetization are type-I with (b) CWW spin rotation and (c) CW spin rotation. (d) A Lorentz micrograph exhibiting the magnetic bubble of $La_{0.825}Sr_{0.175}MnO_3$, obtained at 90 K and $B$ = 570 mT, and the inset showing the in-plane magnetization mapping calculated from TIE [27]. (e) Schematic illustration of a transition process from the magnetic stripe domains to the magnetic bubbles. Magnetization is depicted by the red arrows.

Figure 4 (color online). (a)–(c) Variation in the magnetic textures at 170 K as a function of the strength of the applied magnetic fields in $La_{0.875}Sr_{0.125}MnO_3$. The external magnetic fields are (a) $B$ = 20 mT, (b) 215 mT, and (c) 250 mT, respectively. The inset of (b) shows an enlarged image of the area enclosed by broken lines in (b) in which the scale bar indicates 250 nm. The yellow broken line represents the contour of the elliptical domain. (d)–(f) Variation in the magnetic textures at $B$ = 350 mT as a function of temperature in $La_{0.875}Sr_{0.125}MnO_3$. The Lorentz micrographs are obtained at (d) 100 K, (e) 130 K, and (f) 160 K, respectively.

Figure 5 (color online). Variation of the magnetic domains as a function of temperature under a constant magnetic field of 200 mT. The images were obtained in the (001) crystal plane at (a) 130, (b) 150, and (c) 170 K, respectively. The uniform magnetic stripe domains in (a) change to the bubble structure in (c).  In the inset of (c), the magnified image of the magnetic bubble, indicated by the yellow arrow, is shown.

Figure 6 (color online). The $T$–$B$ phase diagram of $La_{0.875}Sr_{0.125}MnO_3$ constructed using the experimental results. The circles show the experimental points. The blue area shows the region in which the magnetic bubble domains were formed. The green and white areas show the regions in which the magnetic stripe domains and the forced FM state were realized, respectively.

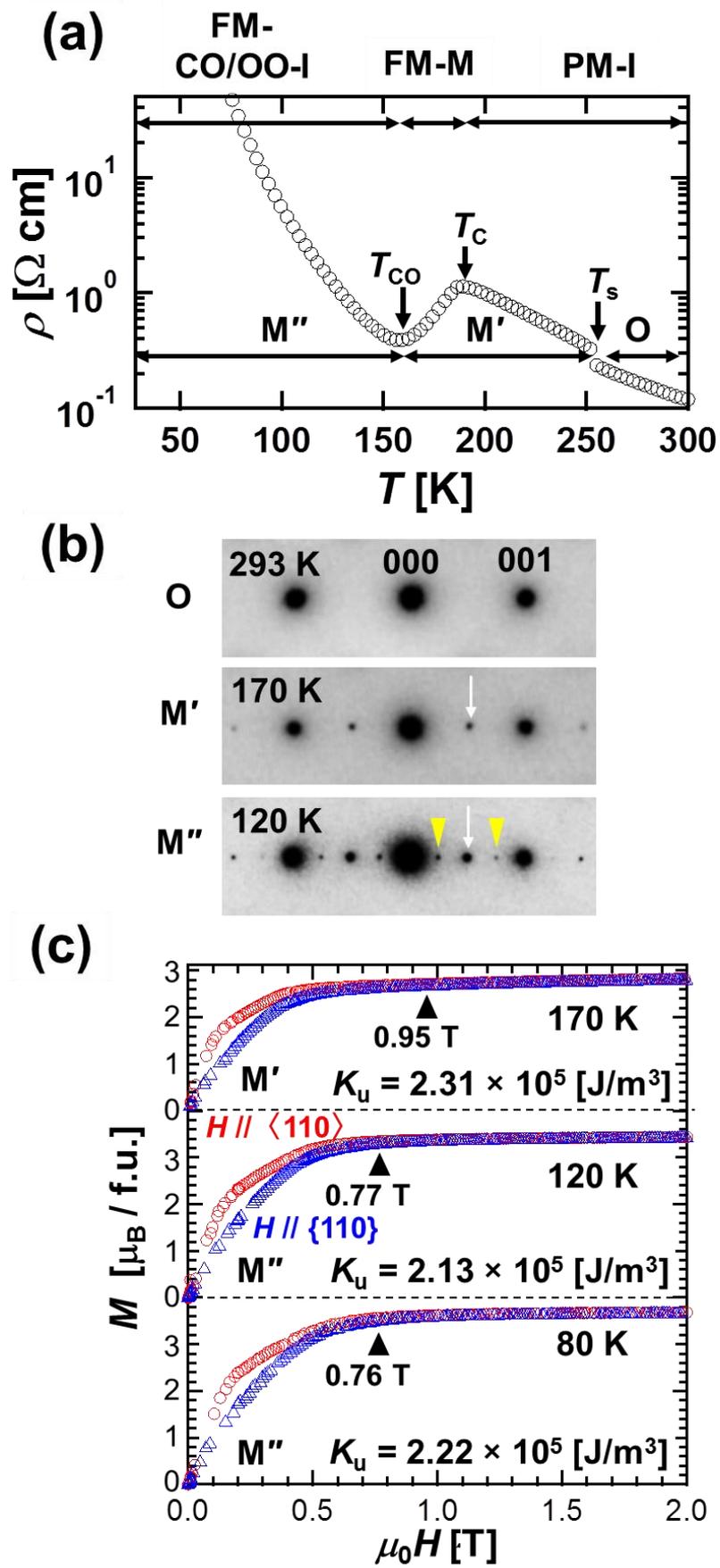

Figure 1, A. Kotani *et al.*, Physical Review B

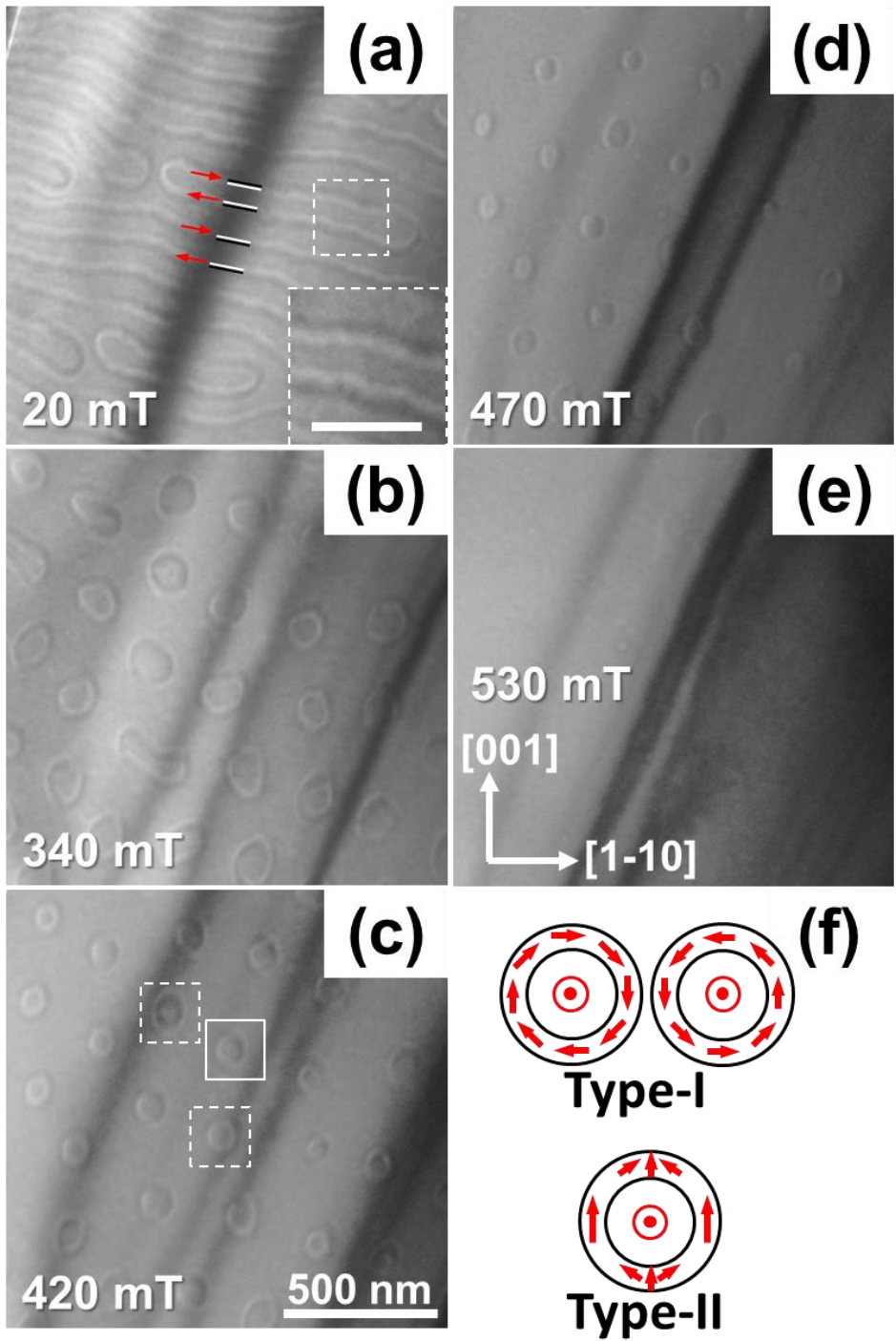

Figure 2, A. Kotani *et al.*, Physical Review B

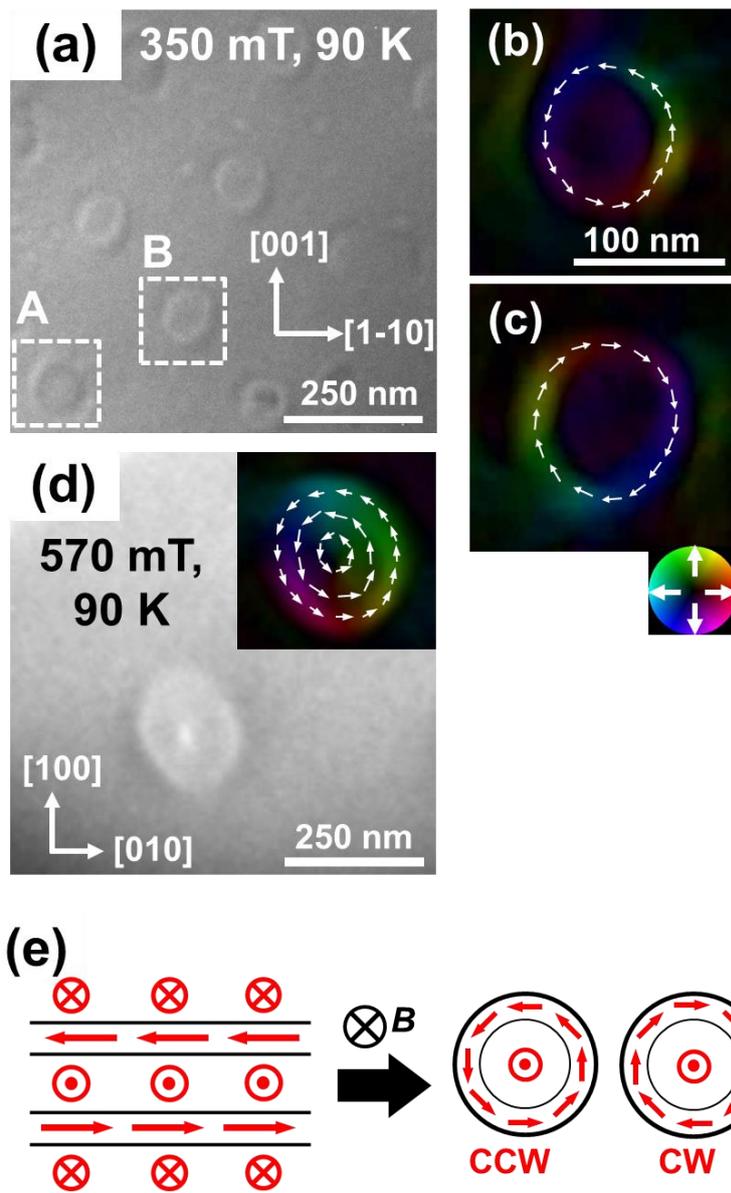

Figure 3, A. Kotani *et al.*, Physical Review B

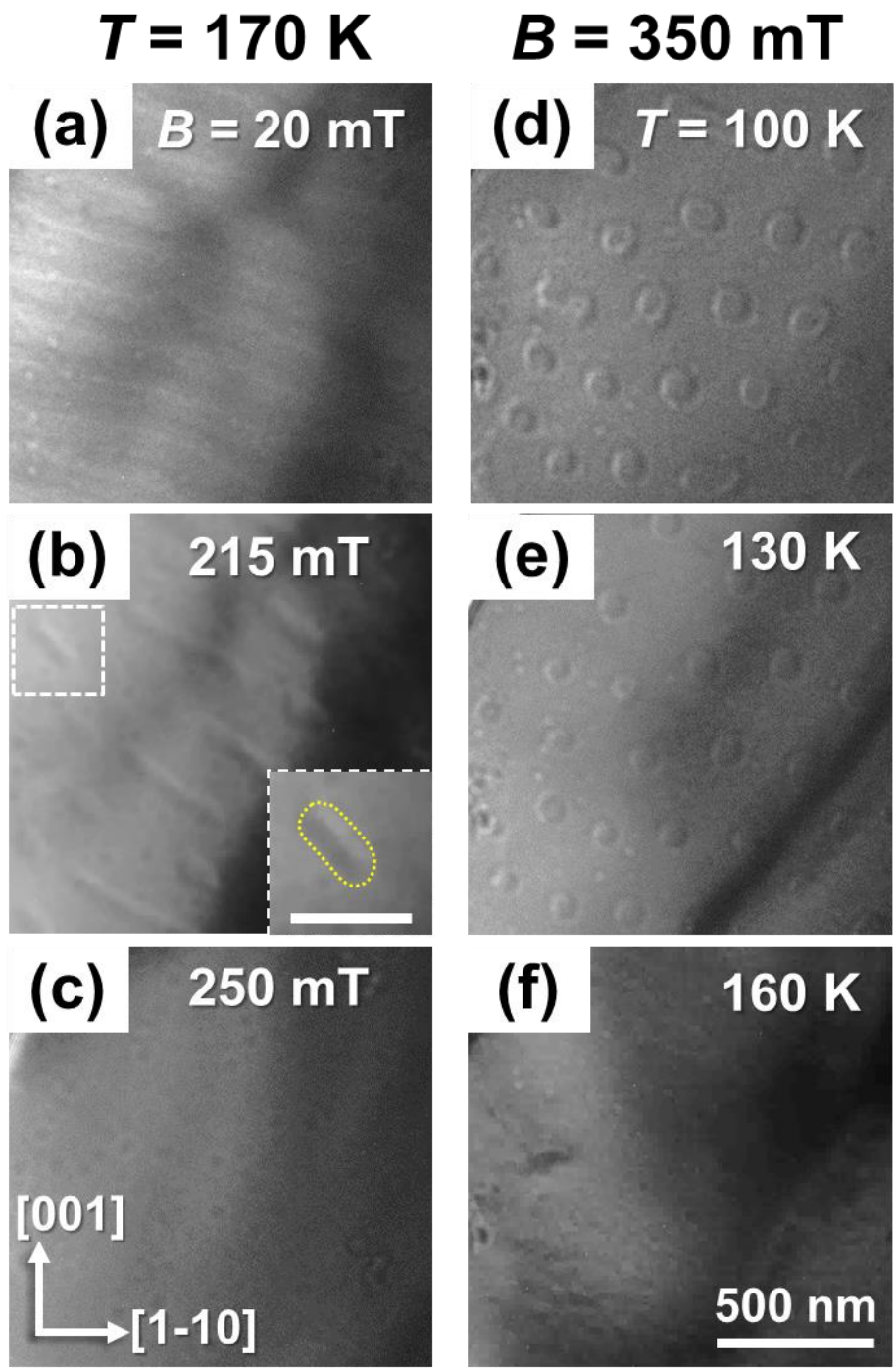

Figure 4, A. Kotani *et al.*, Physical Review B

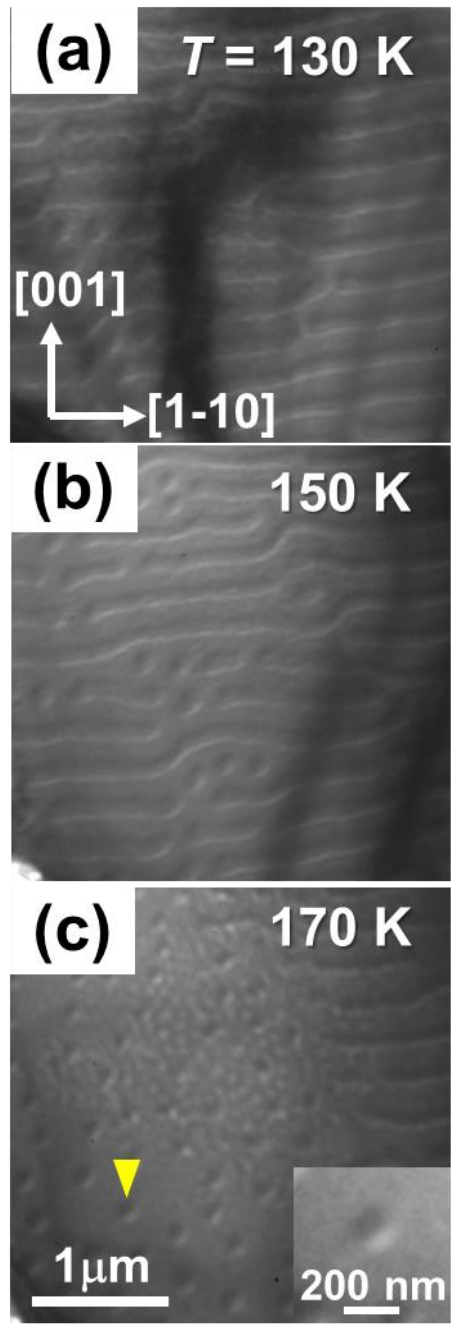

Figure 5, A. Kotani *et al.*, Physical Review B

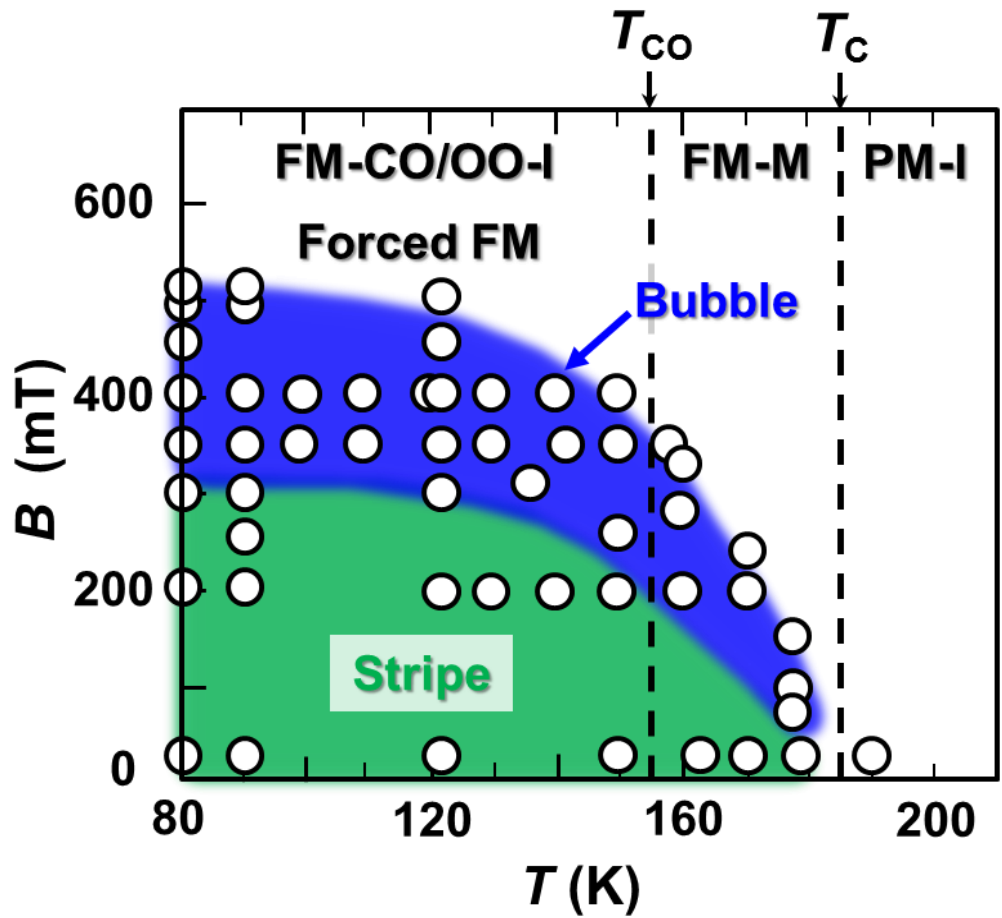

Figure 6, A. Kotani *et al.*, Physical Review B